\documentstyle[11pt,newpasp,twoside,epsf]{article}
\def\spose#1{\hbox to 0pt{#1\hss}}
\def\lesssim{\mathrel{\spose{\lower 3pt\hbox{$\mathchar"218$}}
             \raise 2.0pt\hbox{$\mathchar"13C$}}}
\def\largersim{\mathrel{\spose{\lower 3pt\hbox{$\mathchar"218$}}
             \raise 2.0pt\hbox{$\mathchar"13E$}}}
\def\ref#1{\parindent=0pt\hangindent24pt\hangafter1
           \baselineskip=20pt{#1}\par}

\def\edcomment#1{\iffalse\marginpar{\raggedright\sl#1\/}\else\relax\fi}
\reversemarginpar

\begin{document}

\title{On the Nature of X-Ray Flashes}
\author{R. Mochkovitch, F. Daigne}
\affil{Institut d'Astrophysique de Paris, 75014 Paris, France}
\author{C. Barraud, J.L. Atteia}
\affil{Observatoire Midi-Pyr\'en\'ees, 31400 Toulouse, France}

\begin{abstract}
We have developed a toy model for internal shocks which has been
used to generate a large number of synthetic GRBs in 
order to find in the parameter
space the conditions which can lead to the formation of X-ray flashes. The 
key condition appears to be a small contrast of the Lorentz factor in the
relativistic wind emitted by the central engine. 
\end{abstract}

\section{Introduction}
The recently identified X-ray flashes (Kippen et al, 2002) have non thermal spectra
with $E_{\rm p}< 50$ keV and weak gamma-ray fluxes 
($\lesssim 0.2$ ph.cm$^{-2}$.s$^{-1}$ in the 50 -- 300 keV energy range).
They were discovered by Beppo-SAX and are currently observed
by HETE 2. They can be normal GRBs at very high redshifts (which however
does not seem compatible with their duration distribution similar
to that of long bursts) or viewed off-axis (Yamasaki et al, 2002) but their soft spectra 
can also have an intrinsic origin. We have tested this possibility 
in the context of the internal shock model where it is supposed that
a central engine generates a relativistic wind with a 
non uniform distribution of the Lorentz factor. 
\section{A toy model for internal shocks}
The evolution of this relativistic wind  
should be followed with a fully hydrodynamical
calculation. This can be done (Daigne and Mochkovitch, 2000) but requires
large amounts of computing time which prevents to consider a large 
number of cases and fully explore the parameter space. These detailed
calculations have however shown that a simplified approach where the
wind is represented by many shells which interact by
direct collisions only (all pressure waves being neglected) can also
produce satisfactory results (Kobayashi, Piran and Sari, 1997; Daigne
and Mochkovitch, 1998). Going a step further we have developed 
for this study a toy model where internal shocks are limited to the 
collision of two shells of equal mass $m$. Shell 2 (Lorentz factor 
$\Gamma_2$) is generated a time $\tau$ after shell 1 (Lorentz factor
$\Gamma_1<\Gamma_2$). The average power injected into the wind 
in this two shell
approximation
is given by 
\begin{equation}
{\dot E}={m c^2\over \tau}\,(\Gamma_1+\Gamma_2)={\dot M}{\bar \Gamma}c^2
\end{equation}
where ${\dot M}=2m/\tau$ and ${\bar \Gamma}={1\over 2}(\Gamma_1+\Gamma_2)$
are the average mass loss rate and Lorentz factor.
Shell 2 will catch up with shell 1 at the shock radius
$
r_{\rm s}=2\,c\tau {\Gamma_1^2\Gamma_2^2\over \Gamma_2^2-\Gamma_1^2}
$.
The two shells merge and the energy dissipated in the collision 
$E_{\rm diss}$
is radiated with a characteristic
broken power law spectrum. If the synchrotron process is responsible
for the emission, the peak energy (maximum of $\nu F_{\nu}$) is 
\begin{equation}
E_{\rm p}\simeq E_{\rm syn}
\propto \Gamma_{\rm s}\,B\Gamma_e^2\propto \Gamma_{\rm s}\,
\rho_{\rm s}^x \epsilon_{\rm s}^y
\end{equation}
In the last term $\rho_{\rm s}$ and $\epsilon_{\rm s}$ are respectively
the typical density and dissipated energy (per unit mass) in the comoving
frame of the shocked material. The standard equipartition assumptions
correspond to
$x=1/2$ and $y=5/2$ but we consider below the possibility that $x$ and $y$
can have different values. This may be for example the case if the
equipartition parameters $\alpha_{\rm B}$ and $\alpha_e$ which fix the
fraction of dissipated energy injected into the magnetic field and the
relativistic electrons are not constant but depend on $\rho_{\rm s}$ or/and
$\epsilon_{\rm s}$. 

The physical properties of the shocked material
$\Gamma_{\rm s}$, $\rho_{\rm s}$ and $\epsilon_{\rm s}$ can be 
directly related to
the wind parameters $\tau$, ${\dot E}$, ${\bar \Gamma}$ and 
$\kappa=\Gamma_2/\Gamma_1$ so that eq.(2) can be expressed as
\begin{equation}
E_{\rm p}\propto {{\dot E}^x \varphi_{xy}(\kappa)\over
\tau^{2x}{\bar \Gamma}^{6x-1}}
\end{equation}
where $\varphi_{xy}(\kappa)$ is an increasing function of $\kappa$
for all reasonable values of $x$ and $y$.
In spite of the simplicity of the two shell approximation eq.(3) predicts
a duration-hardness relation and a hardness-intensity correlation (HIC)
as observed in real bursts. Another interesting (and surprising) 
consequence of eq.(3) is that $E_{\rm p}$ 
is a decreasing function of $\bar{\Gamma}$ as long as $x>1/6$.
. Soft bursts are obtained with
``clean fireballs'' (large $\bar \Gamma$) while 
winds with larger baryon loads (``dirty fireballs'') lead to 
a harder spectrum because
internal shocks occur closer to the source at a higher density. 

From the values of $E_{\rm diss}$, $\tau$, $E_{\rm break}$ and assuming a
Band spectrum with low and high energy indices $\alpha=-1$ and $\beta=-2.5$ it 
is possible to estimate (for a given redshift) 
the average flux or count rate in any spectral band. The simplicity of the two 
shell approximation then allows to construct 
a large number of synthetic bursts
to check if XRFs can be formed for some specific choice of the 
wind parameters.
\section{A statistical approach}
The redshift $z$ and the four wind parameters $\tau$, ${\dot E}$, 
${\bar \Gamma}$ and $\kappa$ have been obtained in a statistical way
for each synthetic burst. If long GRBs (and XRFs) are related to the 
explosive death of some special class of massive stars their birth rate 
is directly proportional to the star formation rate $\psi$ and their 
distribution in redshift can be 
deduced from 
$\psi(z)$ (Porciani and Madau, 2001).
The distribution of the observed duration $t_{90}$ for long bursts
is log-normal 
with a maximum at $t_{90}\sim 20$ s. We therefore also take 
a log-normal distribution for
$\tau$ (with a maximum at $\tau_{\rm max}=10$ s)
assuming an average 
burst redshift
$\langle z \rangle\sim 1$. 
The last three wind parameters $\dot E$, $\bar \Gamma$ and $\kappa$
are very poorly constrained and we adopt for them uniform distributions:
between 50 and 53 for ${\rm Log}{\dot E}$, 100 and 500 for $\bar \Gamma$,
0 and 1 for ${\rm Log}\,\kappa$. 

We performed a first numerical experiment with $10^6$ synthetic bursts
to obtain their $E_{\rm p}$ distribution. The results are shown in Fig.1
for two choices of $x$ and $y$: $x=1/2$ and $y=5/2$ i.e. synchrotron 
emission with standard equipartition assumptions and $x=y=1/4$. This last 
case was already considered by Daigne and Mochkovitch (2003) 
who have shown that it leads to 
very good fits of the temporal and spectral properties of 
individual GRB pulses.
The dashed line in Fig.1 represents the distribution of $E_{\rm p}$ 
for all $10^6$ bursts while the full line corresponds to the sub-group  
of bursts which would have been detected by BATSE (a threshold 
of 0.2 ph.cm$^{-2}$.s$^{-1}$ in the 50 -- 300 keV energy band was assumed).
It can be seen that for $x=1/2$ and $y=5/2$ the distribution of $E_{\rm p}$
is much too wide while the agreement with the observations is excellent for
$x=y=1/4$. Figure 1 also shows that a large fraction of events with 
$E_{\rm p}<100$ keV is not detected by BATSE but may contribute to 
the population of XRFs seen by Beppo-SAX and HETE 2.

To test this possibility we have generated (assuming $x=y=1/4$) 1000
events for which we compute both $E_{\rm p}$ and the photon flux in 
the 50 -- 300 keV band. The results are compared to the BATSE + Beppo-SAX
data (Kippen et al, 2002) in Fig.2. Small dots represent BATSE bursts while large dots
are events which are not detected by BATSE but have an X-ray flux 
(2 -- 10 keV) larger than 1 ph.cm$^{-2}$.s$^{-1}$. The similarity between
the two diagrams clearly shows that XRFs can be produced
in the context of the internal shock model.      
\begin{figure}
\plottwo{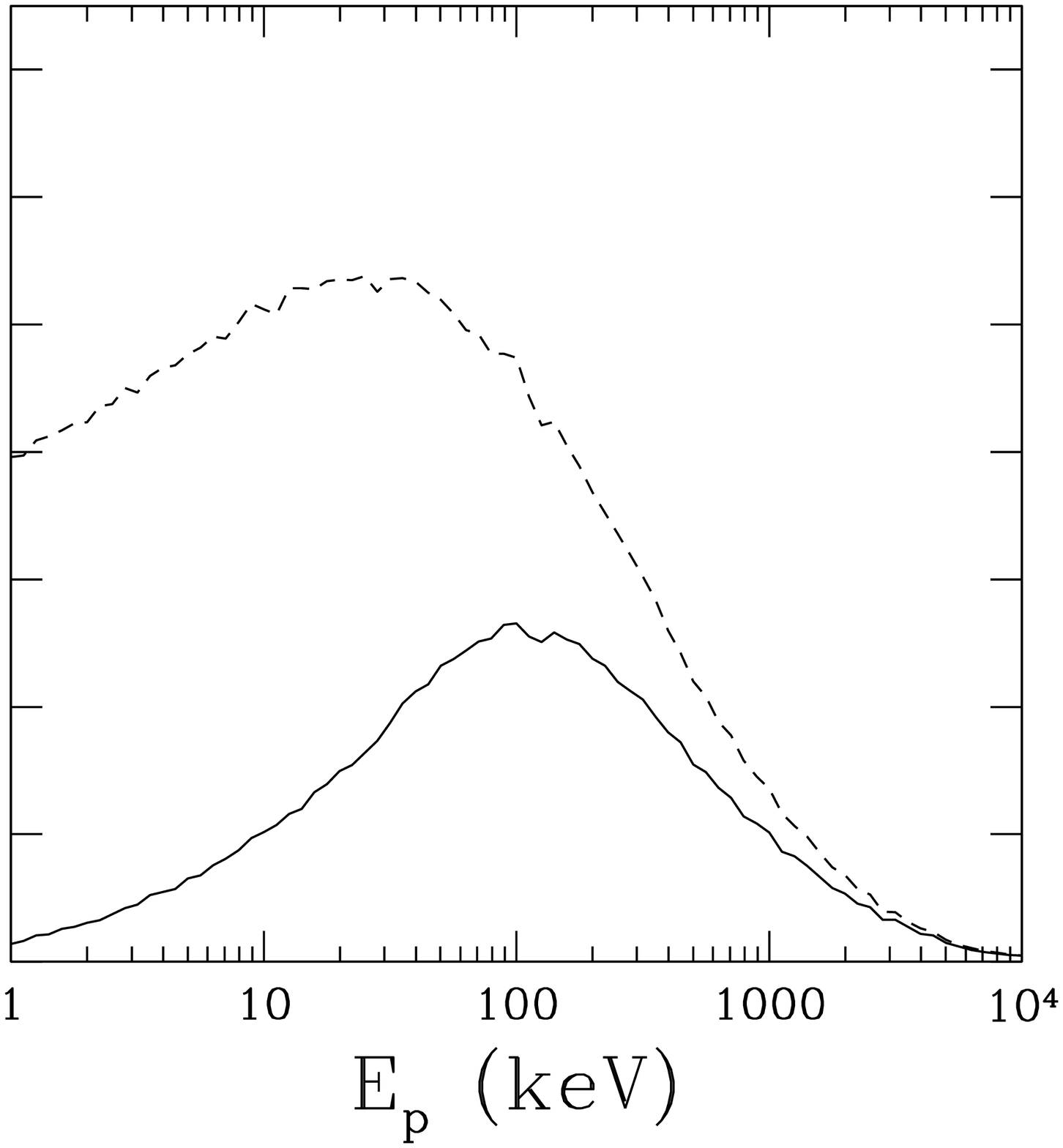}{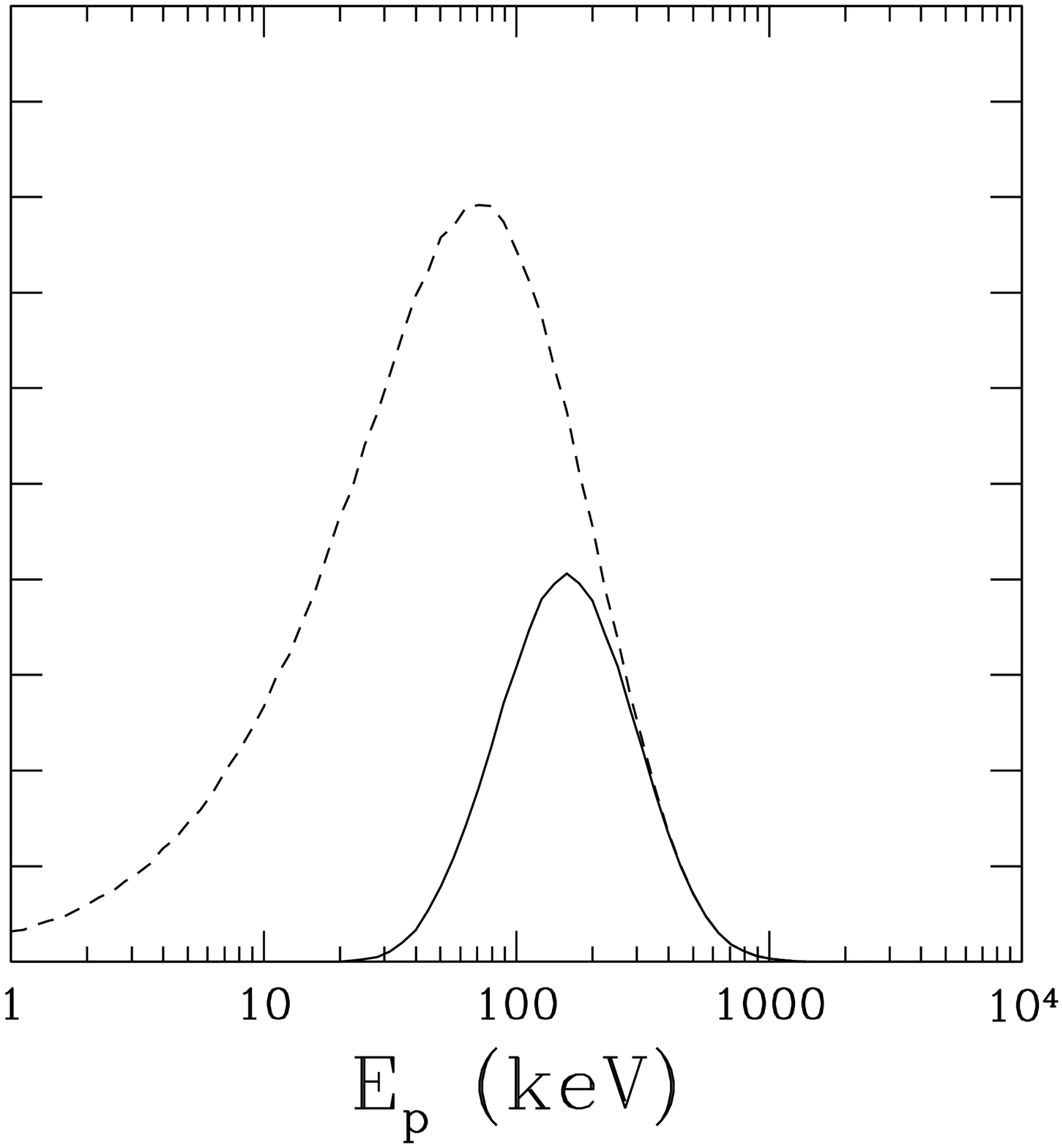}
\caption{Distribution of $E_{\rm p}$ for the synthetic burst population:
left ($x=1/2$ and $y=5/2$); right ($x=y=1/4$). The dahed line represents
the whole population and the full line 
the sub-group of bursts which can be detected by BATSE.}
\end{figure}
\begin{figure}
\plottwo{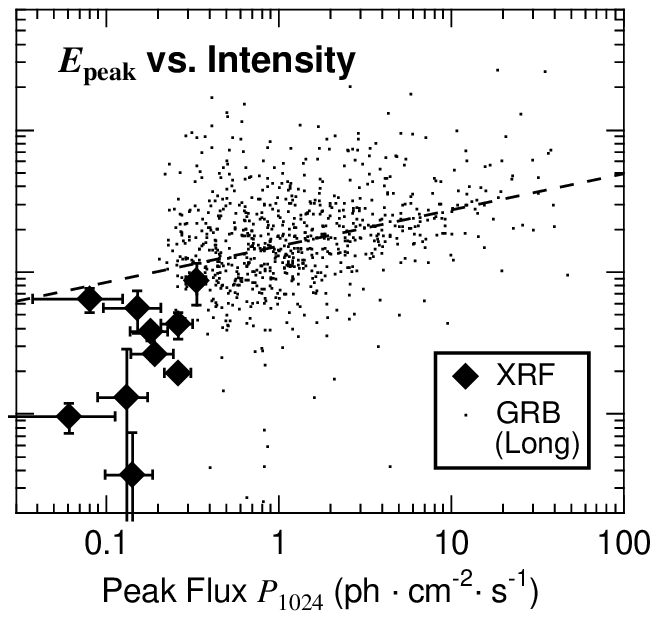}{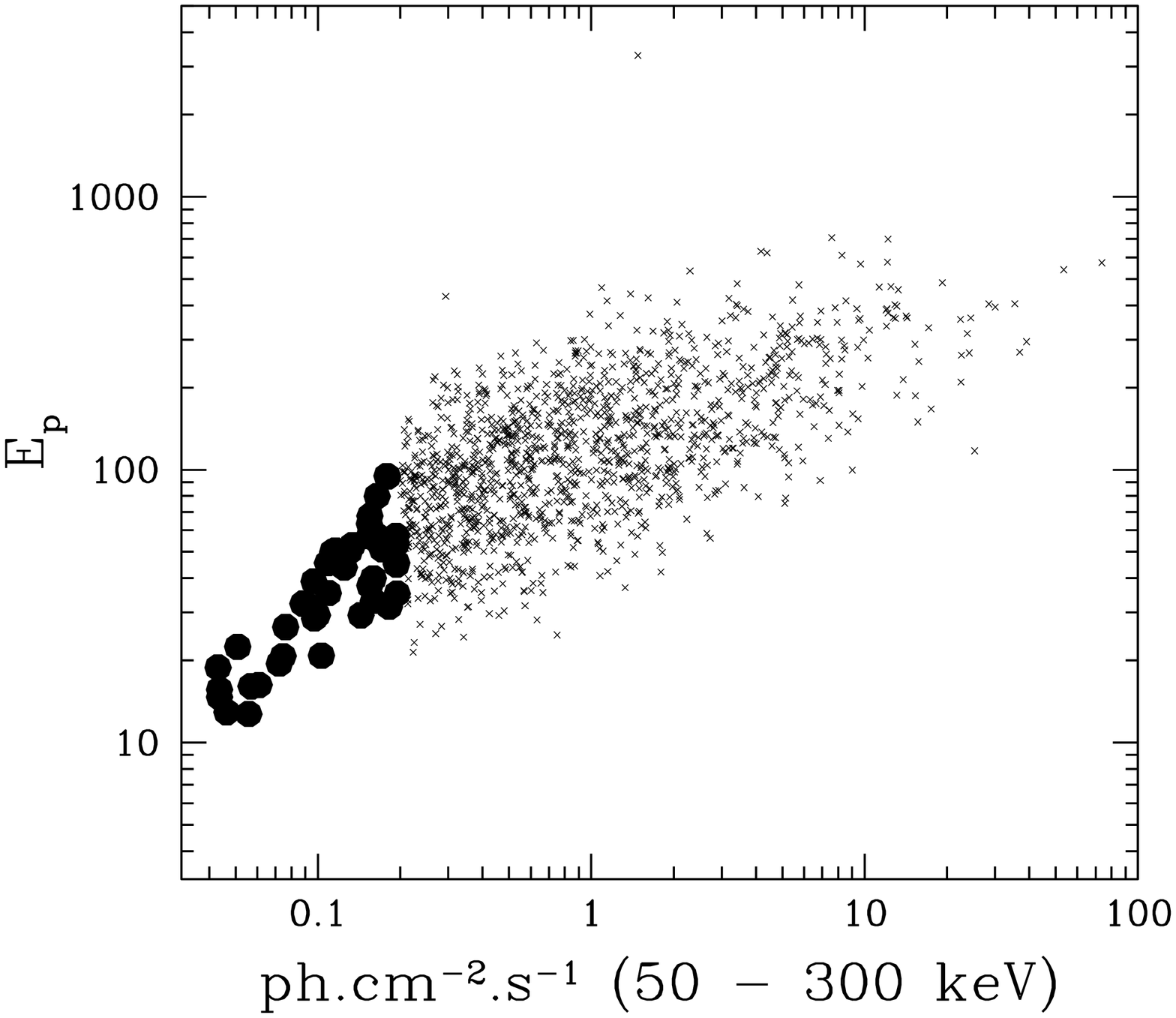}
\caption{$E_{\rm p}$ -- photon flux (50 -- 300 keV) diagram: 
BATSE and Beppo-SAX data from Kippen et al (2002) (left) 
compared the result of our simulation (right).}
\end{figure}

We then obtained the distributions of $z$, $\tau$, ${\dot E}$, 
${\bar \Gamma}$ and $\kappa$ for synthetic XRFs. 
Their redshift distribution is nearly identical to that of GRBs 
and our simulation therefore 
confirms that XRFs are not GRBs at very high redshifts. 
The duration and injected power distributions are also 
not very different between XRFs and GRBs and it finally appears that 
a reduction of the contrast $\kappa=\Gamma_2/\Gamma_1$ 
and an increase of the average Lorentz factor ${\bar \Gamma}$ are
the most efficient ways to produce XRFs. The distribution of $\kappa$ 
has a sharp maximum at about 1.5 in XRFs while it steadily increases
(and is negligible for $\kappa<2$) in GRBs. 
There are also five times more XRFs with $\bar \Gamma=500$ 
than with $\bar \Gamma=100$.
\section{Conclusion}
Our simulations have shown that XRFs can 
be produced by internal shocks 
if the Lorentz factor in the relativistic wind
has variations of small amplitude only and a high average value.
XRFs are not the result of a low injected
power $\dot E$ but are weak and soft mainly due to  
a small contrast of the Lorentz factor which leads
to a reduced efficiency of the dissipation in shocks.  

\end{document}